\documentclass[prb,twocolumn,showpacs,aps]{revtex4}
\usepackage {graphicx}
\pdfoutput=1
\begin{document}
\title{How Gold nanoparticle acquires magnetism? - Formation of large orbital moment at the interface}

\author{S. Banerjee$^a$\footnote{Email:sangam.banerjee@saha.ac.in, See also arXiv:0906.1497v2 for more data on magnetic property}, S. O. Raja$^b$, M. Sardar$^c$, N. Gayathri$^d$, B. Ghosh$^a$, A. Dasgupta$^b$}

\address{$^a$ Surface Physics Division,  Saha Insitute of Nuclear Physics, 1/AF Bidhannagar, Kolkata 700 064, India\\ $^b$ Department of Biochemistry, University of Calcutta, 35 B. C. Road, Kolkata-700019,\\ $^c$ Material Science Division, Indira Gandhi Center for Atomic Research, Kalpakkam 603 102, India,\\ $^d$ Material Science Section, Variable Energy Cyclotron Center, 1/AF Bidhannagar, Kolkata 700 064, India}

\begin{abstract} 
In this paper, we have tried to find out the origin of magnetism in Gold nanoparticles (Au-NPs). We observe that upon incorporating Gold nanoparticles (Au-NPs) in Fe$_3$O$_4$ nanoparticle medium the net magnetisation increases compared to the pure Fe$_3$O$_4$ nanoparticle medium. This increase of magnetization can be attributed to the large orbital magnetic moment formation at the Au/magnetic particle interface indicating that magnetism observed in Au-NPs is an interfacial effect. This interfacial effect has been supported by the observation of sudden transition from positive saturated magnetisation to a negative diamagnetic contribution as a function of magnetic field on citrate coated gold Au-NPs.
\pacs {75.20.-g,75.50.Lk, 75.50.Tt, 75.75.+a}
\end{abstract}

\maketitle

Magnetism in nano/colloidal particles has become a subject of intense research interest in recent years \cite{bedanta,dormann1,dormann2}. Their rich contribution
to fundamental physics and their importance in technological application has become well established now \cite{freitas}. The observation of decrease in diamagnetic susceptibility of copper, silver, gold and even antimony, bismuth and graphite on colloidalisation has been a puzzle since long ago (see the series of papers publised in 1920's and 1930's \cite{vaidhianathan}). Now we believe that we understand the appearance of magnetism in these type of systems as not due to the atomic spins but due to the orbital moments occuring at the defect sites as we have pointed out in the case of ZnO \cite{sangamapl}.

Multicomponent nanoparticles(NP) shows many interesting magnetic\cite{npmag}, optical\cite{npop} and catalytic\cite{npcat} properties. Core-shell NP's with magnetic materials (metallic or insulating) as core and non magnetic (metallic or insulating) materials as shell, are an active field of research to achieve multifunctionality in a single material. Au coated Fe$_3$O$_4$ NP's are an attractive system\cite{gold} that might have interesting magnetic and optical
properties and important biomedical applications because of negligible cytotoxicity of Au.

In a parallel development over the last decade, ferromagnetism in graphite \cite{graphite}, non-magnetic oxides and borides \cite{sangamapl,oxides,borides} have been reported and the ferromagnetic hysteresis observed in these systems have also been attributed to orbital magnetism \cite{sangamapl,simu} occuring due to nanosize defects/structures.  Polymer stabilized metallic NP's like Au and Pd were found to be magnetic\cite{hori}. Thiol capped Au nanoparticles\cite{peng,garta,venta} and even bare Pd, Au clusters made by gas evaporation method were also found to possess finite magnetic moments\cite{tani}. In this paper we have carried out a systematic study to understand the role of gold in modifying the magnetization of Fe$_3$O$_4$ NPs upon incorporation of Au NPs. We have observed that when Au-NPs covers Fe$_3$O$_4$ NPs resulting in a core-shell structure, the magnetism enhances drastically. In any of the works reported earlier, no study has been done as a function of increasing Au NPs size/content. In this paper we report the synthesis, structure and magnetic characterization of  composites of Fe$_3$O$_4$ NPs and Au NPs with increasing particle size/content of Au. The results could be modelled by attributing the magnetism as arising due to the interfacial effect between the Au and the Fe$_3$O$_4$ NPs. To give a supporting evidence to this we have compared the results obtained with that of Au-NPs coated with citrate which estabilshes that the magnetism observed in these composite systems is arising due to the interfacial effect between the Au and the other component which may be either magnetic or non-magnetic.

Fe$_3$O$_4$ NPs were initially prepared by co-precipitation method. 4 gm ferric chloride and 2 gm ferrous chloride (2:1, w/w ratio) were dissolved in 2 M HCl and co-precipitated by 100 ml 1.5 M NaOH solution upon constant stirring for 30 minutes at room temperature. The prepared colloidal solution was centrifuged to collect the supernatant (suspendend) solution to obtain particles with a narrow size distribution. The supernatant solution was pelleted down by a strong magnet and washed four times by ultra pure water. Finally 20 ml Citrate buffer (1.6 gm Citric acid and 0.8 gm tri-sodium citrate) was added to collect the stabilized ferrofluid in solution at a pH around 6.3. This solution was used as a base in the subsequent prepartion of the nanocomposite samples. The solution was lyophilized to obtain the pure Fe$_3$O$_4$ sample which will be referred to subsequently as Sample A. The following procedure was adopted to prepare the Au:Fe$_3$O$_4$  nanocomposite samples: 300$\mu$L of the synthesized colloidal iron oxide nanoparticle (~0.1M) suspension was added to 25ml ultra pure boiling water under vigorous stirring condition. Then 350$\mu$L of 20mM HAuCl$_4$ is added and finally 300$\mu$L of 100mM Tri-sodium citrate was added. The whole solution was kept boiling and stirred for 15 minutes till the color of the solution turned from black to red. The TEM measurement on this sample revealed polydispersed nanoparticles with 5-10 nm size along with very large particles ($\sim$ 200 nm in size) having core-shell structures with Fe$_3$O$_4$ at the core and Au as the outer shell. To seperate out these large particles, the red solution was further centrifuged. In fig~1 we show the TEM micrograph of the (a) supernatant and (b-d) pellet solutions. The particles appearing with lower contrast are Fe$_3$O$_4$ particles and those with high contrast (dark) are the Au particles. In both the samples the Fe$_3$O$_4$ particles are typically 3-4 nm in size. The Au particles in the supernatant sample are nearly monodispered with particle size $\sim$5-6 nm whereas in the pellet sample, they are polydispersed with the particle sizes ranging from $\sim$ 7-10 nm. In the low magnification micrographs fig.1(c-d) of the pellet sample, we could now easily observe the very large particles ($\sim$ 200 nm in size) having core-shell structures with Fe$_3$O$_4$ at the core and Au as the outer shell. These large core-shell particles were not observed in the supernatant sample. The supernatant and the pellet solutions were lyophilized to obtain the dry supernatant sample (Sample B) and the pellet sample (Sample C) respectively. This resulted in two samples with different Au NP's sizes keeping the Fe$_3$O$_4$ particle size same. We would like to mention here that the concentration of the auric chloride taken for our sample preparation seems to be a critical concentration for obtaining these large core-shell particles. We could not obtain these large particles either with a lower or higher concentration of auric chloride. The Atomic Absorption Spectroscopic (Varian AA240) analysis was performed to measure the iron and gold content in the supernatant and in the pellet. The particles were digested in aqua regia (HCl: HNO3 =3:1) to prepare the samples. The standards were 1, 5, 10 and 25 mg/L for both (iron and gold). Aqua regia in same proportion was used as a blank to avoid the iron contribution from HCl and water. In the supernatant the Au and Fe content were found to be 0.46mg/L and 3.789mg/L respectively. Hence, percentage of gold was 10.82\% in the supernatant sample. In the pellet sample the Au and Fe content were found to be 5.132mg/L and 12.749mg/L respectively giving a percentage of gold to be 28.7\%.  TEM images and the corresponding electron diffraction obtained from Fe$_3$O$_4$ particles and the core of the high Au content core-shell particles are shown in Fig.~2. The electron diffraction rings obtained from the samples could be indexed to that of Fe$_3$O$_4$. In fig.~2(d) we can see that there is a clear signature of Fe$_3$O$_4$ from the core of the Au-Fe$_3$O$_4$ core-shell structure. The density of Fe$_3$O$_4$  is much lower than that of Au and hence we observe more transmission through the centre of the particle indicating that the core region is mainly Fe$_3$O$_4$. The Au-NPs coated with citrate were prepared by a similar procedure. 350$\mu$L 20mM Auric Chloride (HAuCl$_3$) solution is added to boiling 25ml milli-Q water under vigorous stirring. Then 100mM 300$\mu$L tri-sodium citrate is added. After sometime (~4mins) a blue colour appears. This colour changes to pink then to light red and finally to deep red. The solution is allowed to stand for another 15mins after the appearance of deep red colour and then lyophilized to obtain the dry sample. The hydrodynamic diameter of the Au-NPs were obtained by dynamic light scattering method (Photon Correlation Spectroscopy) and was found to be about 22 nm.
The magnetic property of all the samples were measured using MPMS-7 (Quantum Design). 

\begin{figure}
\includegraphics*[width=7cm,angle=270]{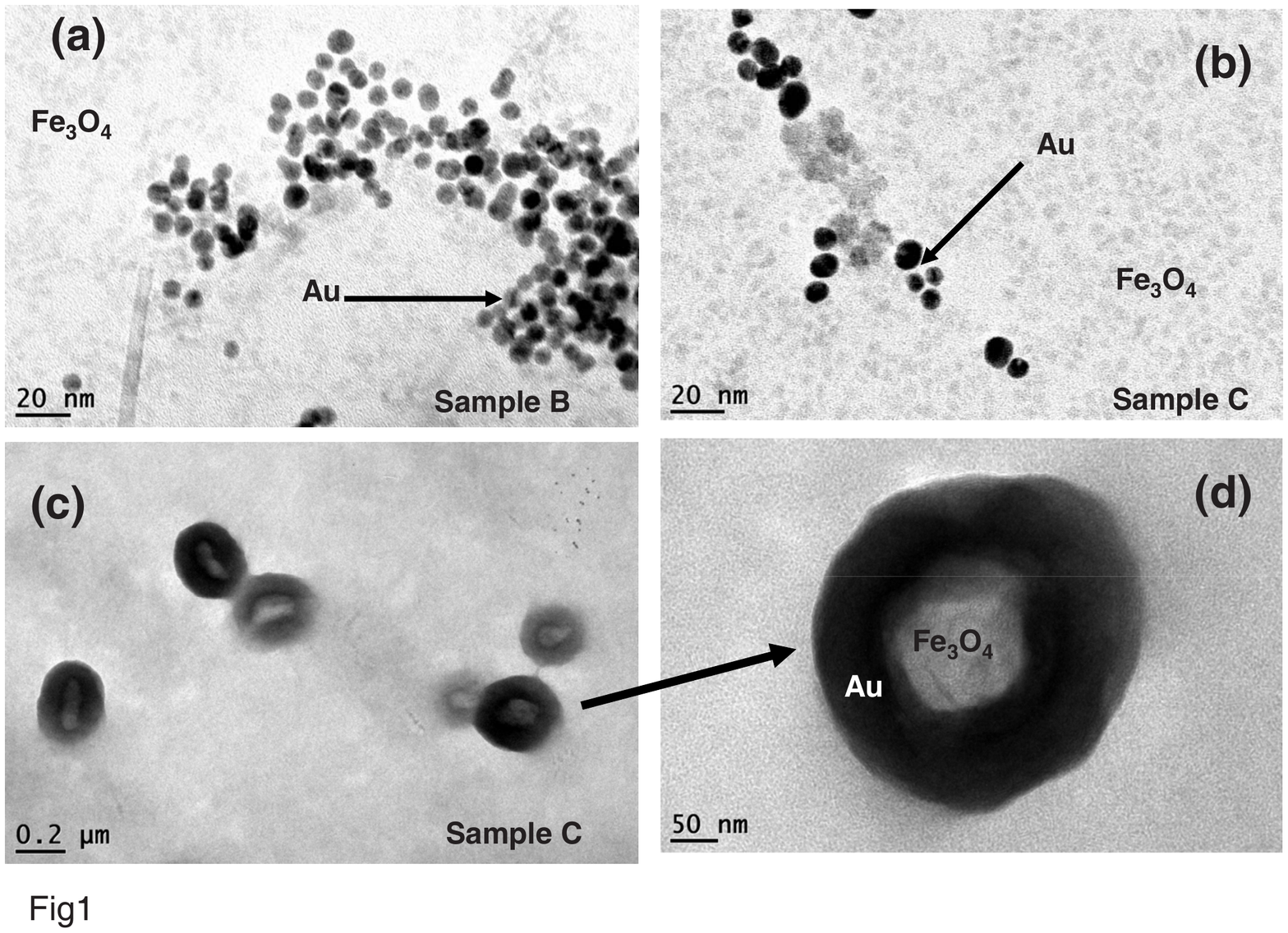}
\caption{(See online for better contrast of Fe$_3$O$_4$ particles) Transmission Electron Micrographs of the (a) Low-Au (Sample B) and (b-d) High-Au (Sample C) nanocomposites. Fe$_3$O$_4$ can be seen in the background as faint particles of size 3-4 nm. Au particles are darker and are marked by arrows. Core(Fe$_3$O$_4$)-shell(Au) structures seen in Sample C are shown in (c) and (d).}
\end{figure}

\begin{figure}
\includegraphics*[width=7cm]{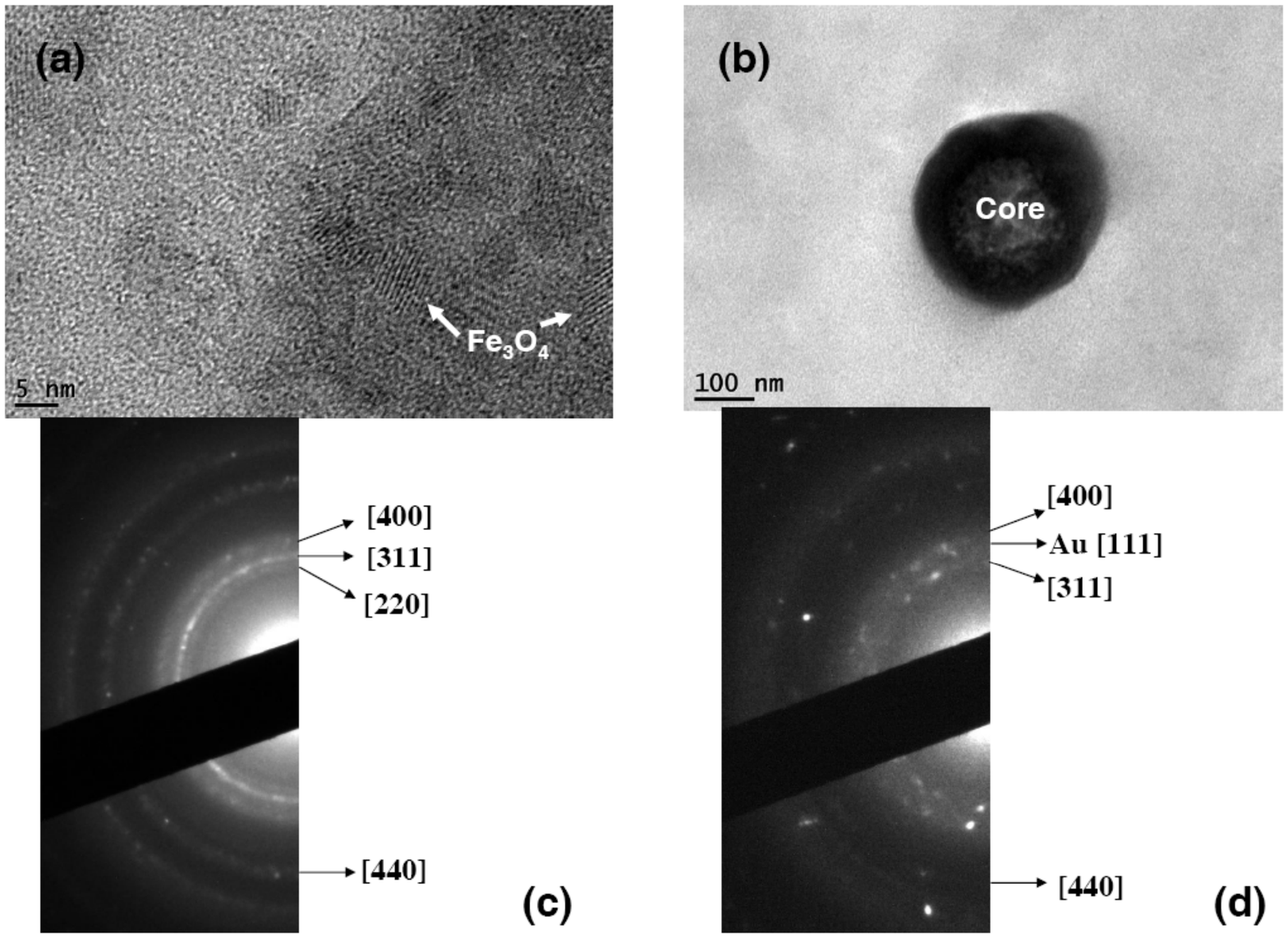}
\caption{Transmission Electron Micrographs and electron diffraction of the Fe$_3$O$_4$ (a) and (c) and the core of the high-Au core shell particles (b) and (d). The diffraction rings have been indexed to Fe$_3$O$_4$.}
\end{figure}

\begin{figure}
\includegraphics*[width=7cm]{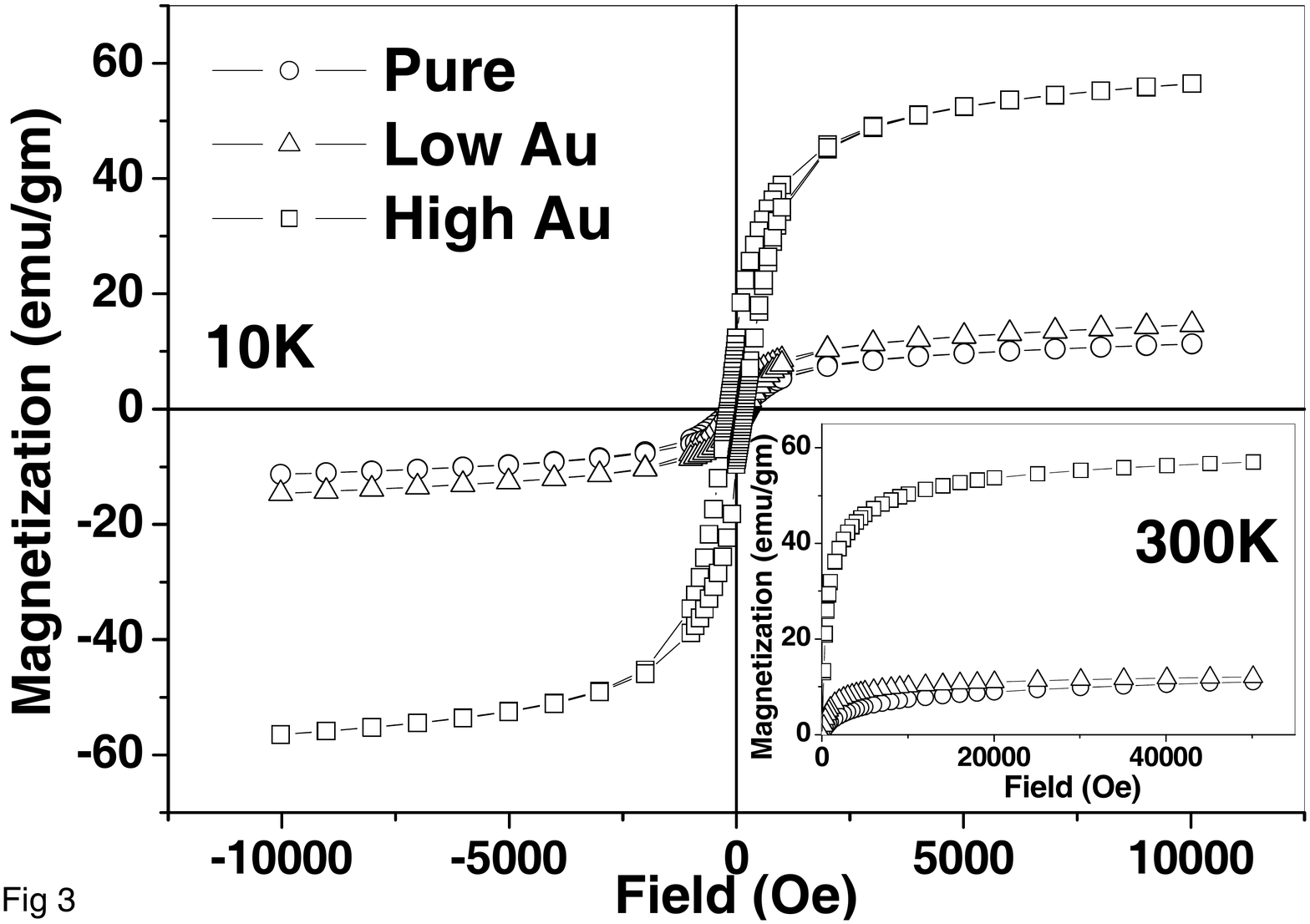}
\caption{Magnetization vs H for Sample A (Pure), Sample B (Low-Au) and Sample C (High-Au) taken at 10K and inset:300K}
\end{figure}

In fig.~3 we show the magnetic hysteresis of the samples taken at 10 K and 300K. We can clearly see that for both the temperatures, the sample B (low Au content) shows a very small enhancement of saturation magnetisation whereas the sample C (high Au content) shows a drastic enhancement in the saturation magnetization. We would like to find out the reason for the anomalous increase in magnetization with increase in Au particle size and content. As we have emphasized before, the Fe$_3$O$_4$ nanoparticles in our sample have a very narrow size distribution as seen from the TEM images. The difference between the three samples is the variation in Au content. The interesting thing is that with increase in Au content the magnetic moment/gm is increasing. This behaviour we believe is new and unexpected. Typically Fe$_3$O$_4$ nanoparticle moment density is much lower than that of bulk Fe$_3$O$_4$ \cite{chen2}(84 emu/gm). This is generally thought to be due to finite size effect and surface spin canting due to lower coordination number and strain or structural deformation at the surface of th NP\cite{surf}. It has been observed that in Au coated  Fe$_3$O$_4$ nanoparticles the moment reduces further\cite{srikanth1,wang}, indicating that surface moments might be further disordered due to interaction with Au electrons, leading to the reduced moment. Our system Au-Fe$_3$O$_4$  sample C is very different from their core-shell structure because their particle sizes are much smaller than our sample. The sample C has large core-shell particles having large interface between Au and Fe$_3$O$_4$.  In any case the disordering of canted moments on Fe$_3$O$_4$ surfaces due to conduction electrons of Au should still be occurring and hence an increase of net moment is rather surprising. A simple guess would be that Fe$_3$O$_4$ is spin polarising Au very close to the interface. This will increase the net moment as well as increase the effective volume of the magnetic nanoparticles. Spin polarization of nonmagnetic metals in contact with ferromagnets was studied extensively by Hauser \cite{hauser} experimentally and theoretically by Clogston \cite{clog}. They found that the spin polarization can at best penetrate a length scale of 1-2 nm in a nonmagnetic metal in contact with a ferromagnet. This is rather small to explain the large change in saturation magnetization coming from an effective volume increase of  magnetic nanoparticle due to spin polarization of Au electrons
 near the interface. Since spin polarization does not extend to large distances, it cannot explain the continued increase in net moment with increase in Au content. Experimentally magnetic moment of Au near Co/Au interface has been measured from magnetic X-ray circular dichroism \cite{coau} to be about 0.062 $\mu_B$ per Au atom near the interface. The origin is the spin-orbit splitting of Au surface states (inversion symmetry is lost on the surface), but the moments are far too small (by a factor of 100) to explain our observed increase in moment in Au- Fe$_3$O$_4$ system. So we have to look elsewhere to explain this phenomena.

A set of interesting experimental results on the magnetic properties of some nanostructures has been recently published. Large magnetic moments were detected on the surface layers of thin films of borides and oxides\cite{oxides,borides}. Ferromagnetic hysteresis at room temperature was
 measured in Au nanoparticles\cite{Au} and Au nanoparticles/films with thiol patches on top\cite{authiol,carmeli}. Spin splitting of surface electronic states was observed in Au(111)\cite{lashell}, and Bi\cite{koroteev}. Similar magnetism was detected in Pd nanoparticles
also\cite{sampedro}. A common characteristics of all of these unusual magnetic behavior seems to be that local anisotropy is very large compared to typical anisotropy strengths of  well known harder materials. Usual understanding of magnetism in polymer or thiol stabilized Au NP's is that, there is considerable amount of charge transfer from Au to polymer or thiol, exposing d-holes,
which are for some reason polarised to give a net moment. Recently an alternative theoretical attempt was made by Hernando et.al.\cite{hernando} to explain magnetic moment in Au with thiol patches on top. Important difference of our system is that here we have an interface between Au and magnetic Fe$_3$O$_4$ unlike Au and non-magnetic thiol. As we shall see, this has significant consequences. 

We shall assume the existence of a contact potential $V$ and a radial electric field
(perpendicular to the interface) $E=-(dV/dr)_{r=\eta}$ at the Au - Fe$_3$O$_4$
interface. This will induce a Rashba type spin-orbit interaction, $H_{spin-orbit}= {\mu_B \over c^2} (v \times E)\cdot s= -\alpha \hbar^2 L_zs_z$. Here $\eta$ is the radius of the interface of any Au particle with the surrounding Fe$_3$O$_4$ particles, $\alpha$ is the spin-orbit
coupling strength and is proportional to the gradient of the contact potential. Free electrons on the surface of the Au nanoparticles, can be captured in large atomic like bound orbitals of radius $\eta$ at the domain boundary potential step. With the spin component of the bound Au electron along the $z$ axis being $s_z$, the Hamiltonian (in the absence of external magnetic field), for
these bound electrons can be written down as 

\begin{equation}
H= \frac{\hbar^2 L_z^2}{2 m \eta^2} -\alpha \hbar^2 L_zs_z  + \lambda s \cdot \sum_{i \in \hbox{interface}} S_i
\end{equation}

\noindent $\lambda$ is the exchange (antiferromagnetic\cite{clog}) coupling strength of the
Au electron having spin $s$ and any Fe moment at the boundary, $i$ being site index of the Fe moments along the interface having a spin $S_i$. The first term is the kinetic energy of the electrons near the interface with angular momentum $L$. The second term is the spin orbit interaction induced by the interface potential gradient. The third term is the contact exchange interaction of these electrons near the interface with the Fe moments on the surface of the Fe$_3$O$_4$ nanoparticles.  We have neglected the Zeeman term proportional to external magnetic field , because it is very small (see the estimate later in the text).

We take the average Au nanoparticle radius as $\eta= 4$ nm (the average size (diameter) of the Au NPs is 7-10 nm as seen from TEM measurements). For the spin-orbit coupling we take, $\alpha\hbar^2 = 0.4$ eV.  This value is large and close to the atomic spin orbit coupling of the 6p states of Au atoms of 0.47 eV\cite{moore}. Experimentally observed spin splitting of the surface states of pure Au(111) surfaces\cite{lashell} could be explained by assuming $\alpha \hbar^2=0.4$ eV. Theoretically it was shown\cite{petersen} with a simple tight binding model for the surface states, that indeed the spin orbit coupling for the surface states can be as large as the atomic spin orbit coupling. It was also pointed out\cite{petersen} that the magnitude of the surface potential (closely related to the work function of Au), as well as other potential steps on the surface, can further increase the effective spin-orbit coupling of the surface states. In our case with interface with oxide particles, we believe that 0.4 eV for the spin orbit coupling interaction is an underestimate.

We also take, $\lambda =0.05 $ eV (typical values of contact exchange interaction)\cite{clog} for illustrative purposes and write the Hamiltonian as

$$
H= {\hbar^2 L_z^2 \over 2 m \eta^2} -\alpha \hbar^2 L_zs_z  + \lambda s_z <M_z> + 
$$
\begin{equation}
\lambda \sum_{i \in \hbox{interface}} {1\over 2}(S_i^+s^- +S_i^-s^+)
\end{equation}

\noindent where $<M_z>=\sum_{i \in \hbox{interface}}<S_{i,z}>$ is the net average z
component moment of the surface Fe atoms. The last part is the transverse part of the contact exchange interaction, that gives
rise to spin flip scattering between the boundary Fe moments and the Au electrons (both bound and free electrons). Forgetting the last term for the time being, we find that when $M_z=0$ the energy is negative for $L_z$ = 1 to 93, i.e, one could have 93 electrons filling such bound orbitals all with same $s_z$. In fig.~4 we have plotted the energy versus $L_z$ values for different values of $M_z$. We can see, that for $M_z >190$ there are no bound states (negative energy) at all, for the chosen values of parameters. In other words if the $z$ component of the boundary spins add upto large values then it is not possible to have bound Au electrons along the interface with large orbital angular momenta. On the other hand when average $M_z=0$ like in Au-thiol (nonmagnetic) interface, it is possible to have large number of bound states occupied with electrons (having $L_z$ values from 1 to large values, and same $s_z$ to minimise exchange part of the coulomb correlation energy) near the interface, giving a large net moment. Interestingly it was found by Crespo et al\cite{crespo2} that with addition of Fe impurity the thiol capped Au nanoparticle magnetism disappears very quickly. This curious observation is easily understandable from the above discussion.

The spin flip scattering (the last term in eqn.2) by the free as well as bound Au electrons with the boundary Fe moments, on the other hand try to randomize the  Fe moments giving rise to lesser $M_z$ value. Thus, samples with larger Au concentration will have larger concentration of free electrons, and hence reduces the average boundary Fe moments more efficiently compared to sample with lesser Au/free electron concentration. This could be one of  the reason why, we find larger moment in samples with larger concentration of Au. If we had for example, a composite of Fe$_3$O$_4$ and any nonmagnetic insulating
particles, then our mechanism does not allow the existence of large orbital magnetic moments. Since the additional magnetic moments come from orbital moments, this implies a high magnetic anisotropy of the Fe$_3$O$_4$ + Au bound electrons composites. It has to be emphasized that the dominant reasons for expecting low values for $M_z$, which is essential for survival of these
interface states is surface spin canting due to lower coordination number and strain or structural
deformation at the surface. Moreover at lower temperatures the surface spins are often in a frozen spin glass state.

Though many groups have worked with Fe$_3$O$_4$-Au nanoparticle composites, to our
knowledge there has been no report so far on such large enhancement of net magnetization of the composite. There could be several reasons for that. (1) Since that $z$ axis should be very well defined throughout the Au-magnetic particle interface, the value of effective $\eta$ is  very small for very small sized Au nanoparticles, or for interfaces where the plane enclosed by $\eta$ deviates from a plane too much. For example in thiol capped Au nanoparticle
system the moment/Au atom is very large in thin films compared to small
nanoparticles\cite{authiol}. (2) On the other hand if the Au particles are large,
then the core diamagnetism of Au electrons may cancel out the large orbital moments
at the interface. Thus, there seems to be an optimum size of the Au particle which will show maximum magnetization, beyond which the diamagnetic term will dominate. In our case the base material (Fe$_3$O$_4$) itself is magnetic and Au-diamagnetism is very small compared to the magnetic base material. 

When $M_z=0$, the net energy of an electron at the interface  with orbital angular momentum $L_z$
and spin $S_z$ is, 

\begin{equation}
E={\hbar^2 L_z^2 \over 2 m \eta^2} -\alpha \hbar^2 L_z S_z
-\mu_B H \cdot
(L_z + g_s S_z)
\end{equation}

\noindent where $g_s=2$. We see that the energy remains negative (bound state) upto,
$L_z^{Max}=\alpha m \eta^2$ (taking $S_z=1/2$). Taking the values of the parameters as in eqn.~2, we find $L_z^{Max}=93$. If we put electrons in orbitals with $L_z=1,2,3....N$ all with $S_z=1/2$ then net moment $M$ is $[\frac{N(N+1)}{2} +N ]$ $\mu_B$. Putting $N=93$ we get $M=4464 \mu_B$.
\noindent The net energy of $N$ electrons is given by, 

$$
E_N= {\hbar^2 \over 2 m \eta^2 }\sum_{n=1}^N n^2 -{\alpha \hbar^2
\over 2} \sum_{n=1}^N n - H \times 4464 \mu_B 
$$
\begin{equation}
={\hbar^2 \over 2 m \eta^2 }{N(N+1)(2N+1)\over 6} -
{\alpha \hbar^2
\over 2} {N(N+1)\over 2}- H \times 4464 \mu_B
\end{equation}

\noindent with $N=93$ the contribution from the first two terms is $E_N=-295.89$ eV. The contribution of the
last term (Zeeman) is only -1.39 eV for a magnetic field of 5 Tesla and hence can be neglected (and this is why we have not considered 
this term in eqn.~1). 

We have neglected the coulomb interaction between these electrons so far. Let us consider it now.
In atomic orbitals of extent $r=1-2 \times 10^{-8}$ cm, the coulomb correlation energy between two electrons in two different orbitals is
about $1-2 eV$; we take 2 eV to have an upper limit on the coulomb repulsion energy. So the average electron-electron interaction energy in orbitals of
size $\eta= 40 \times 10^{-8}$ cm is about ${2\over 40}=0.05 eV$. Total coulomb interaction energy of N electrons is about,
$E_{coulomb}=+{N(N-1)\over 2}\times 0.05 = + 213.9  eV$. So we see that the total energy $E_{tot}=E_N +E_{coulomb}$ is still
negative for $N=93$, indicating that it is possible to have many electrons at the interface. 

Now we are ready to make an estimate about how much moment one should
expect in sample B and Sample C. From the measured high field magnetisation we find that for pure
iron oxide particles (sample A), the magnetisation is 12 emu/gm. Taking a density of 5 gm/cm$^3$ , we have $6 \times 10^{21} \mu_B$/cm$^3$ moment for pure Fe$_3$O$_4$.

The high field magnetisation of the sample B is about 14 emu/gm
or about $7 \times 10^{21} \mu_B$/cm$^3$ , assuming the same
density. With 10.82\% Au in sample B, the net moment/cm$^3$ coming
from 89.18\% Fe$_3$O$_4$ and orbital moment from the interfaces of 10.82\% Au is about,

$$
M_{Total}=0.8918 \times 6 \times 10^{21} \mu_B + { 0.1082 \over
  {4\pi \over 3} \times \eta^3} \times 4464 \mu_B 
$$
\begin{equation}
= 7.154 \times 10^{21} \mu_B
\end{equation}

This is close to the high field magnetisation $7 \times 10^{21} \mu_B$ of sample B
Now assuming 28.7\% Au in sample C, we find that the expected net magnetisation/cm$^3$ should be about $9.06 \times 10^{21} \mu_B $/cm$^3$. The measured high field magnetisation of sample C, on the other hand is about 58 emu/gm, or $29 \times 10^{21} \mu_B$/cm$^3$, which is considerably higher than the calculated value. This large value of the magnetization can be accounted by considering a very large interfacial area of the large core-shell structured Fe$_3$O$_4$-Au particles in Sample C. As we have mentioned before we see a few large
sized core-shell type of particles of diameter 200 nm range. Since, $L_z^{Max} \propto \eta^2$, the net orbital magnetic moment per Au NP $M={L_z^{Max}(L_z^{Max}+1)\over 2} \propto (L_z^{Max})^2 \propto \eta^4$. Thus the large particles (core-shell particles) with interface radius about 50 nm ie., the radius of the core Fe$_3$O$_4$ particle will have orbital magnetic moments about $10^4$ times larger than the smaller NP's. In fig.~5 we have plotted the magnetic moment at the interface of a Au-Fe$_3$O$_4$ NP versus the interface radius of the NP, showing the steep increase of the moment with increase in size of the interface.

The difference between the calculated and the observed value of the moment for sample C is around $20 \times 10^{21} \mu_B$. This additional moment can be obtained by simply considering just 2\% of the total particles to be core-shell structure with $\eta = 50 nm$ and M = 10$^4$ times the moment of the smaller NP as mentioned above. Hence we infer that the huge increase in the magnetic moment of Fe$_3$O$_4$-Au composites in sample C is coming from mainly the large size core-shell NP's.

\begin{figure}
\includegraphics*[width=7cm]{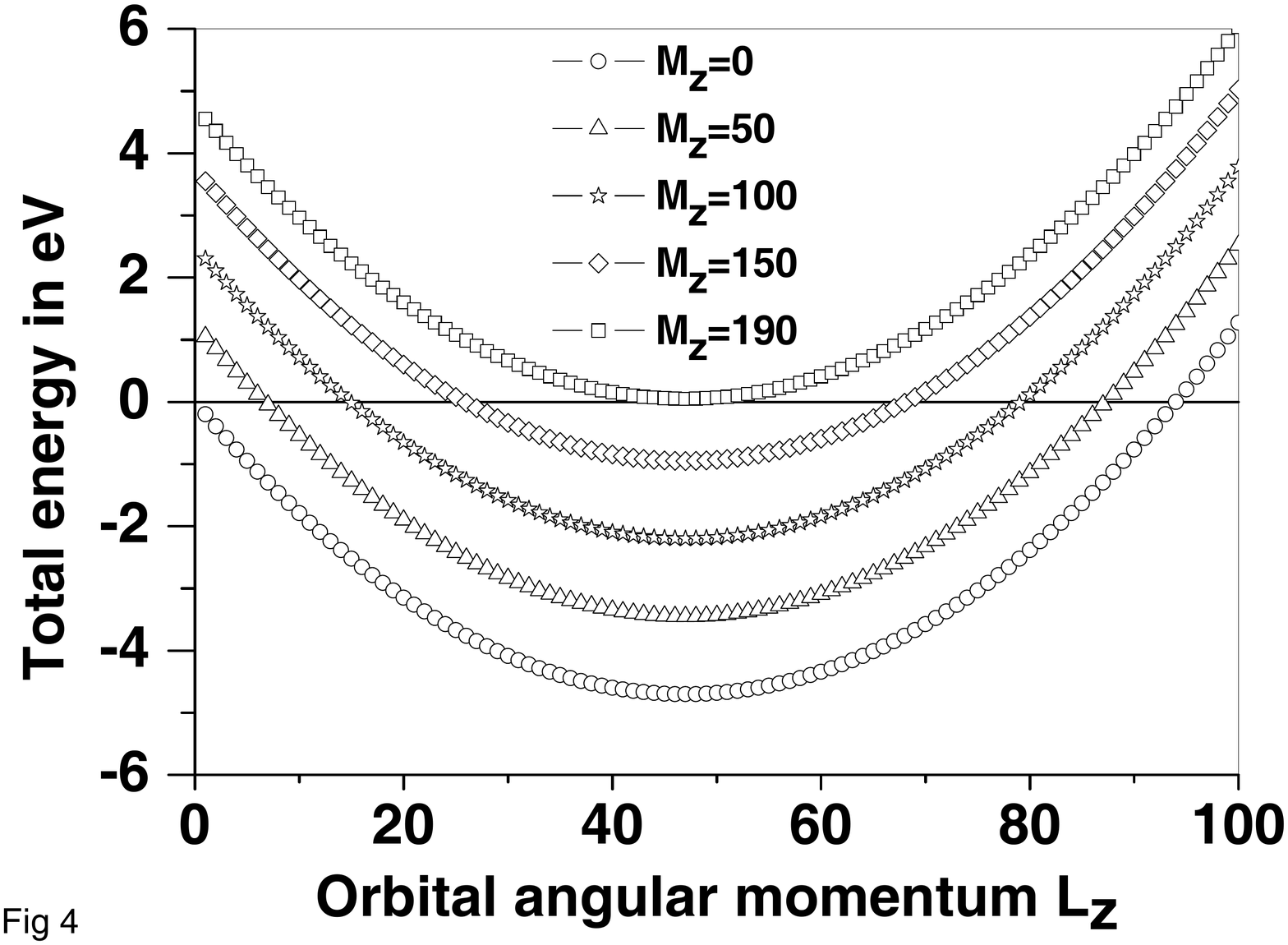}
\caption{Total energy in eV versus angular momentum $L_z$ of an electron near the interface.}
\end{figure}

\begin{figure}
\includegraphics*[width=7cm]{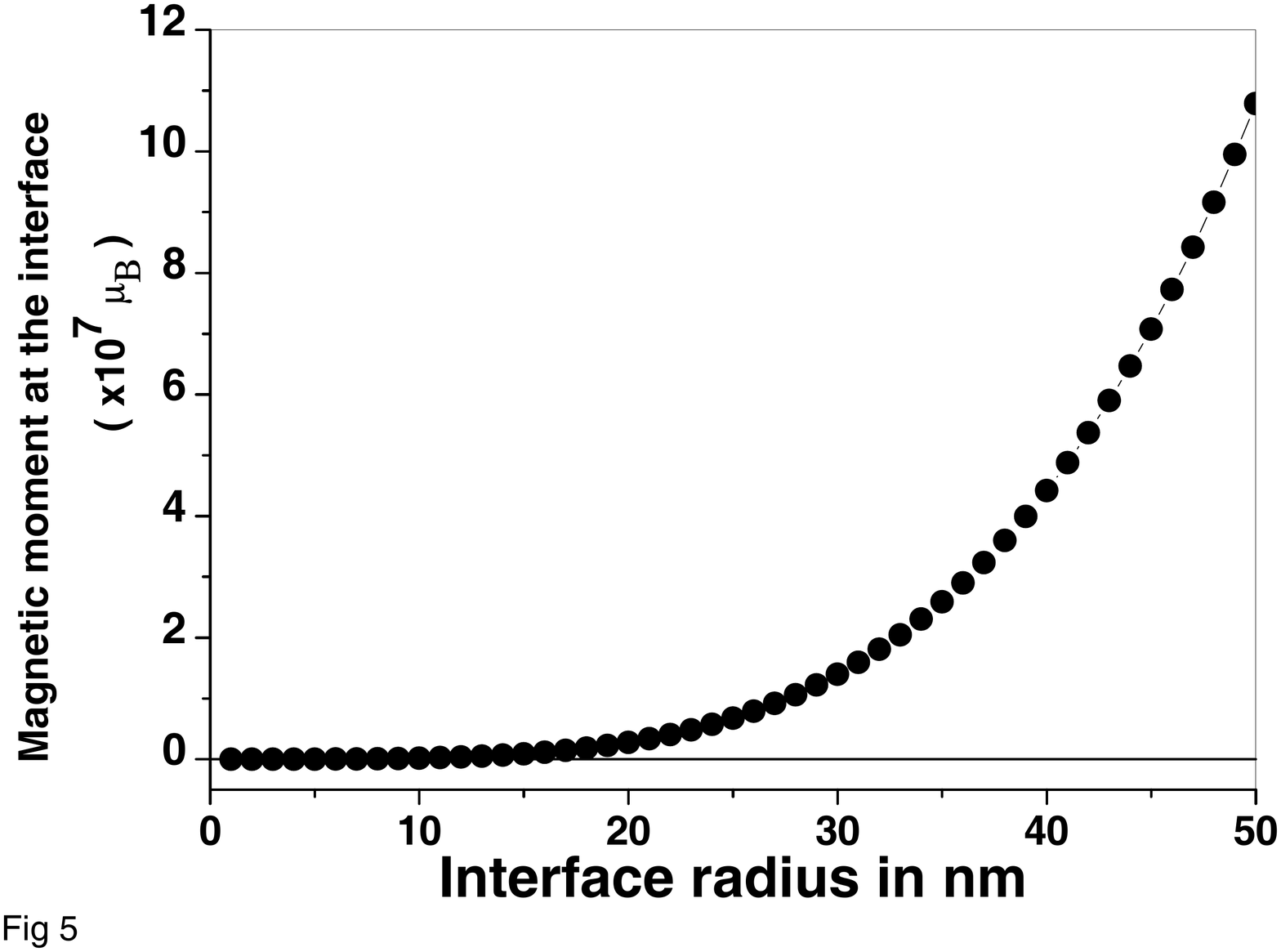}
\caption{Net Magnetic moment (orbital+spin) of the bound
electrons near the interface(in $\mu_B$) versus the radius of the interface (in nm).}
\end{figure}

\begin{figure}
\includegraphics*[width=7cm]{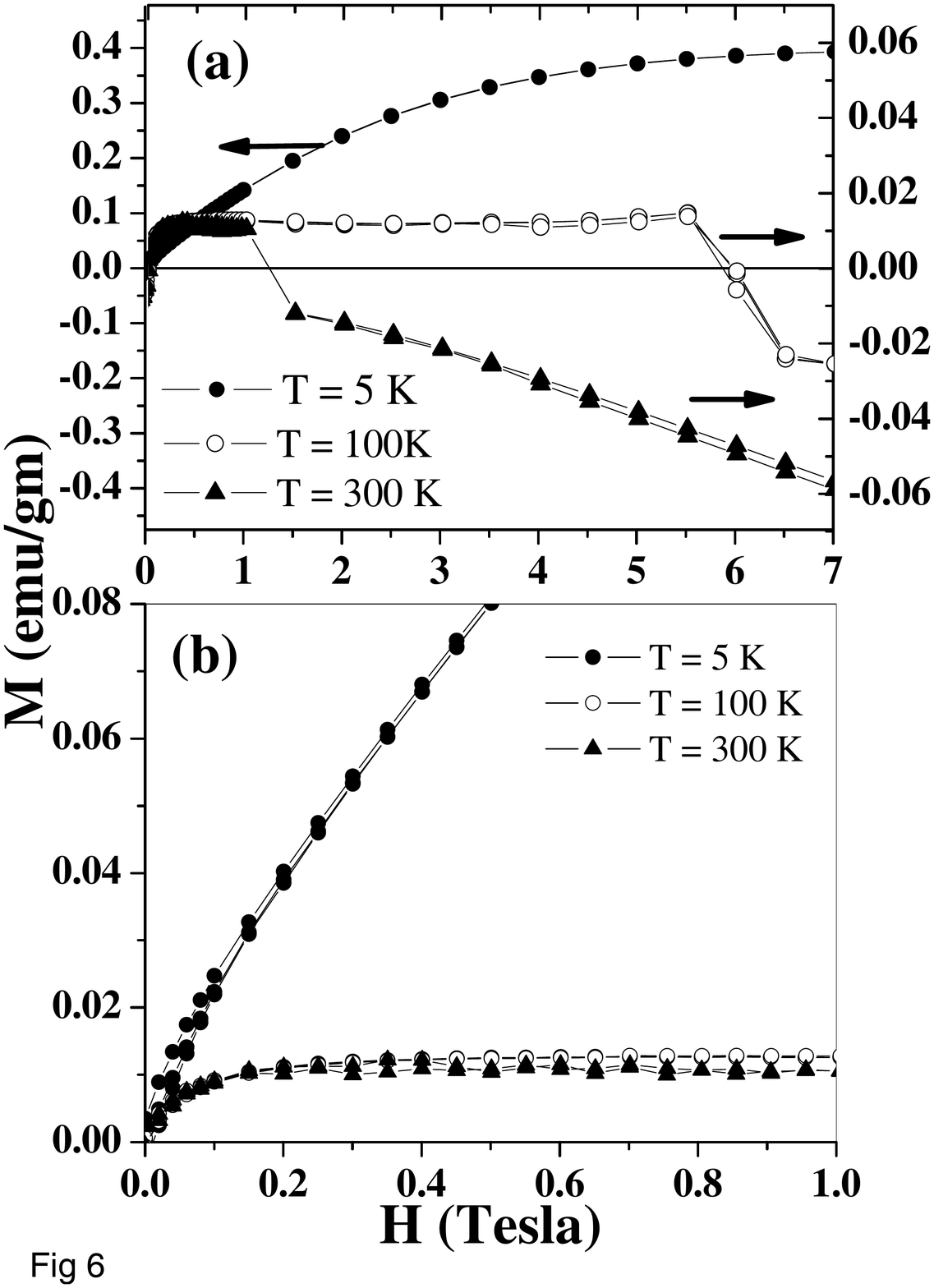}
\caption{(a) Magnetization vs H for the Au-citrate sample taken at 5K, 100K and 300K upto a field of (a) 7 Tesla and (b) expanded scale upto 1 Tesla. }
\end{figure}

If the magnetism in Au NPs arise due to the interfacial contact potential, then this should also be seen in a system containing interface of Au with non-magnetic material. We have used Au-NPs coated with citrate to verify magnetism arising due to the interfacial effect. Fig.~6 shows the M vs H curves of the Au-citrate NP's taken at three different temperatures (T = 5K, 100K and 300K) upto 7 Tesla magnetic field. We observe the following. (1) The saturation magnetisation at 5K (occuring beyond 7 Tesla) is an order of magnitude more than that at 100K and 300K. The saturation at the higher temperture occurs at around 0.2 Tesla (see fig.~6(b)). (2) The magnetisation at 100K and 300K shows a sudden transition from a magnetic to a diamagnetic state at $\sim$ 6 Tesla and $\sim$ 1 Tesla respectively. (3) Faint hysteresis is observed at 5 K. 

With an average Au NP radius of 11 nm, and the saturation magnetic moment at the lowest tempertures $T=5$ K, and a field of 7 Tesla, we estimate
the average magnetic moment/Au NP to be about $1115 \mu_B$. If $L_m$ is the maximum $L_z$ value of any bound state electron, then the net
moment of that particular Au NP is about 

\begin{equation}
M={L_m(L_m+1)\over 2} + L_m \approx {L_m^2 \over 2}
\end{equation}

\noindent so we get , ${L_m^2 \over2}=1115 \mu_B$ ,or, $L_m=47 \mu_B$. Since, also, $L_m= \alpha m \eta^2$, we find that  the necessary spin-orbit coupling, to get $L_m=47$ is about, $\alpha \hbar^2$= 56.76 meV.

Remembering that for Fe$_3$O$_4$-Au composite we have taken $\alpha \hbar^2=0.4$ eV, the value for Au-citrate interface is less
than 0.4 eV by a factor of 7. This is reasonable, as can be seen from the following arguments.  We have assumed that, in an interface between metal/insulator
there is charge transfer from the metal to the insulator. But a good upper bound of the amount of charge transfer, will be
when the field created at the interface due the charge transfer is nearly equal to the dielectric breakdown field of the insulator.  The  dielectric  breakdown field of organic polymers is about, 50-900 KV/cm. Thin films of metal oxides dielectric breakdown field depends on
defect concentration as well as thickness. For SiO$_2$ for example, this
field varies from 1-10 MV/cm \cite{badila,sa} and for organic thin films like Polyethilines, benzene, this field is about 0.5 MV/cm \cite{david}. It is likely that transition material oxide like Fe$_3$O$_4$ will  presumably have a breakdown field of the same order of magnitude as SiO$_2$.  Since spin-orbit coupling is proportional to this field, a factor of 7 seems reasonable. 

Using the arguments given earlier, we can explain the sudden transition from the magnetic to a diamagnetic state observed in the magnetisation at higher temperatures. Let us assume that the total magnetisation observed in these particles is due to the addition of the diamagnetic and the interfacial ferromagnetic contributions. The interfacial moment contribution at 5K is very large compared to that at 100K and 300K because we observe the saturation magnetisation to be much smaller at 100K and 300K (observation (1) above). Since diamagnetic contribution is independent of temperature, at higher tempertures, the diamagnetic can overcome the low interfacial magnetic contribution at a lower field (observation (2) above). Hence we see a sudden change from a low (positive) saturation magnetisation to a (negative) diamagnetic behaviour. Assuming a paramagnetic scaling of the orbital moment ($M\propto {H/T}$), and assuming the bulk diamagnetic susceptibilty (temperature independent)
 of Au to be about, $\chi_{dia}=2.8 \times 10^{-6}$ emu/cc/Oe, we find the field at which the total magnetic moment should do the zero crossing 
for tempertures,  T= 100 K and T=300 K should be about, H=4 Tesla and H=0.6 Tesla. Experimentally these fields are 5.5 Tesla and 1 Tesla respectively. Note: The sudden change in the magnetisation is an interesting observation since one would have expected a gradual change from a positive to a negative value. This issue is beyond the scope of this work and will be addressed in a future publication. But however, this observation clearly indicates that the interfacial effect is the most likely phenomena for the observed magnetism in Au NPs.

In the present investigation we have observed that it is possible to increase the net magnetic moment in Fe$_3$O$_4$-Au nanocomposites by increasing the Au content. The chemical potential gradient at the interface of the Fe$_3$O$_4$-Au is enough to trap conduction electron from the Au particle and induce large orbital moment at the interface. Thus the enhanced magnetic moment is argued to come from metallic electrons at the Au-Fe$_3$O$_4$ interface and predominantly orbital in origin. We have also found that to have very large increase in net magnetic moments it is necesary to have core-shell type particles with large interface. We have done quantitative estimates of such induced magnetic moments and compared with our experimentally measured value. The agreement is reasonably good. We have also shown that this interfacial effect is the most likely phenomena for the observation of magnetism in Au-NPs by showing magnetism in citrate coated gold Au-NPs. The observation of sudden transition of positive magnetisation to negative diamagnetic magnetisation as a function of magnetic field could be explained by the destruction of the interfacial magnetic moment by the diamagnetic contribution.

{\bf Acknowledgement} The authors thank the TEM facility at Institute of Physics, Bhubaneswar for the TEM measurements.

\end{document}